\documentclass[superscriptaddress,aps,pra,twocolumn]{revtex4}

\usepackage{graphicx}
\usepackage{dcolumn}
\usepackage{bm}

\usepackage{bm}
\usepackage{ulem}
\usepackage{epsfig}
\usepackage{graphicx}
\usepackage{amssymb,amsmath,amsbsy,amsgen,amsfonts}    
\usepackage{dcolumn}
\usepackage{amsthm}
\usepackage{mathrsfs}
\usepackage{latexsym}
\usepackage{array}
\usepackage{color}
\usepackage{amstext}
\allowdisplaybreaks[1]
\usepackage{txfonts}
\usepackage{pifont}
\usepackage{braket}
\usepackage{epstopdf} 

\usepackage[dvipsnames]{xcolor}
\usepackage[hidelinks,colorlinks=true, linkcolor=WildStrawberry, urlcolor=WildStrawberry, citecolor=Emerald, linktocpage=true, breaklinks=true, bookmarksopen=true]{hyperref}

\newcommand{\abs}[1]{|#1|}

\newcommand{\be}{\begin{equation}}
\newcommand{\ee}{\end{equation}}
\newcommand{\ba}{\begin{array}}
\newcommand{\ea}{\end{array}}
\newcommand{\bqa}{\begin{eqnarray}}
\newcommand{\eqa}{\end{eqnarray}}

\DeclareSymbolFont{symbols}{OMS}{cmsy}{m}{n}

\begin{document}

\title[]{Verification of single-photon path entanglement using a nitrogen vacancy center}
\author{A. I. Smith}
\affiliation{Department of Physics, Stellenbosch University, Private Bag X1, Matieland 7602, South Africa}
\author{C. M. Steenkamp}
\affiliation{Department of Physics, Stellenbosch University, Private Bag X1, Matieland 7602, South Africa}
\affiliation{National Institute for Theoretical and Computational Sciences (NITheCS) South Africa}
\author{M. S. Tame}
\email{markstame@gmail.com}
\affiliation{Department of Physics, Stellenbosch University, Private Bag X1, Matieland 7602, South Africa}
\affiliation{National Institute for Theoretical and Computational Sciences (NITheCS) South Africa}
 
\date{\today}

\begin{abstract}
Path entanglement is an essential resource for photonic quantum information processing, including in quantum computing, quantum communication and quantum sensing. In this work, we experimentally study the generation and verification of bipartite path-entangled states using single photons produced by a nitrogen-vacancy center within a nanodiamond. We perform a range of measurements to characterize the photons being generated and verify the presence of path entanglement. The experiment is performed using continuous-wave laser excitation and a novel state generation `time-window' method. This approach to path entanglement verification is different to previous work as it does not make use of a pulsed laser excitation source.
\end{abstract}
 
\maketitle


\section{Introduction}
Quantum entanglement plays a fundamental role in many quantum information protocols~\cite{Nil10}. In quantum communication, for example, entanglement is used in quantum key distribution~\cite{Ekert91}, quantum dense coding~\cite{Bennett92,Bruss04,Yeo06,Mattle96}, quantum secret sharing~\cite{HBB99,CGL99,Tittel01,Bell14} and quantum teleportation of unknown quantum states~\cite{Bennett93,Ren17}, amongst many other protocols~\cite{Gisin02,Gisin07,Scarini09,Pers2013}. While entanglement has been generated in a wide range of physical setups~\cite{Hor09}, photonic systems are particularly well suited for generating entanglement and using it in quantum communication~\cite{Gisin02,Gisin07}, quantum computing~\cite{Cou23} and quantum sensing~\cite{Degen17,Cou23b}. The first step in many quantum photonic applications is to generate photons that can then be entangled using additional optical components. Over the years, trapped single atoms or ions have traditionally been considered as a photon source~\cite{Cou23}. However, solid state emitter systems in the form of nitrogen vacancy (NV) centers within diamond have become leading candidates for the generation of single photons, due to their atom-like properties, with a rich energy-level structure and stable solid--state setting~\cite{Aha11,Child13}. 

In this work, we perform an experiment that uses an NV center within a nanodiamond as a single-photon source and explore its ability to generate single-photon path entanglement by extending a previously introduced method~\cite{Lou09,Pap09}. The previous experiment of Papp {\it et al.}~\cite{Pap09} used an ensemble of cesium atoms contained within a magneto-optical trap and pulsed laser excitation, making the system challenging to realise. In our work, we show that a stable room temperature NV center within a nanodiamond can be used as the single-photon source and continuous wave (CW) laser excitation is enough to generate path entanglement, making the method simpler and more accessible. Single-photon path-entangled states may be used for quantum communication~\cite{Cas20,Bjork12, Scarini09,Cou23}, as resources in very-long-baseline interferometry for astronomy and geodesy~\cite{Brown23,Rajagopal24,Wang24}, and for applications in quantum computing~\cite{Cou23} and quantum sensing~\cite{Cou23b}.

Quantum entanglement has two alternative definitions: the first, more mathematical one, states that a composite system which forms a state is entangled if it is not separable~\cite{Hor09}. The second definition is more physically motivated and states that entangled states are states that cannot be simulated by classical correlations~\cite{Mas07}. In our work, entanglement is generated across two paths into which a single photon is sent~\cite{van06}. The state of the system at its simplest is $\ket{\Psi} = \frac{1}{\sqrt{2}}\left(\ket{0}_1 \ket{1}_2 + \ket{1}_1 \ket{0}_2 \right)$, where $\ket{0}_i\ (\ket{1}_i)$ represents zero (one) photon being present in path $i$. In terms of the first definition of entanglement, this state is considered entangled if the states of the two paths cannot be separated, {\it i.e.} if a photon is present in one path then no photon can be present in the other and vice versa. It is important to note that it is the path modes of the photon that are entangled in this scenario and although only involving one photon, the state can in principle be converted into a more traditional entangled state with two photons through local operations, thereby demonstrating entanglement in the original state~\cite{van06, van06a,Chou05}. According to the second definition, the state $\ket{\Psi} $ is considered entangled if it cannot be replicated through classical means, {\it i.e.} it cannot be replicated through the use of classical light being sent down the two paths. We show that both definitions of entanglement are satisfied in our experiment through the use of specialized measurements that consider both the wave and particle nature of the photon.

The NV center that we use as a single-photon source is a defect within a nanodiamond lattice where a missing carbon atom (vacancy) is present, with a substitutional nitrogen atom as one of its nearest neighbors~\cite{Aha11,Ber15}. Two forms of these defects are most commonly found in the literature and used for quantum applications. The first is the neutral state, NV$^0$, which has five unpaired electrons, four from neighboring carbon atoms and one from the nitrogen. The second is the negatively charged state, NV$^-$, which has an additional electron~\cite{Aha11}. Although the two defects are very similar in the way they form, there are a few differences, most notably the wavelength of photons produced when there is no phonon broadening, known as the zero-phonon line (ZPL). For the NV$^0$ the ZPL is $575\ \mbox{nm}$ and for the NV$^-$ it is $637\ \mbox{nm}$~\cite{Aha11}. We chose a single NV$^0$ for our work on generating path entanglement, however, the method of entanglement generation extends to NV$^-$ emitters, as well as other promising solid-state emitters, such as quantum dots~\cite{Lodahl15,Chatterjee21}.
\begin{figure*}[t]
    \includegraphics[width= 17cm]{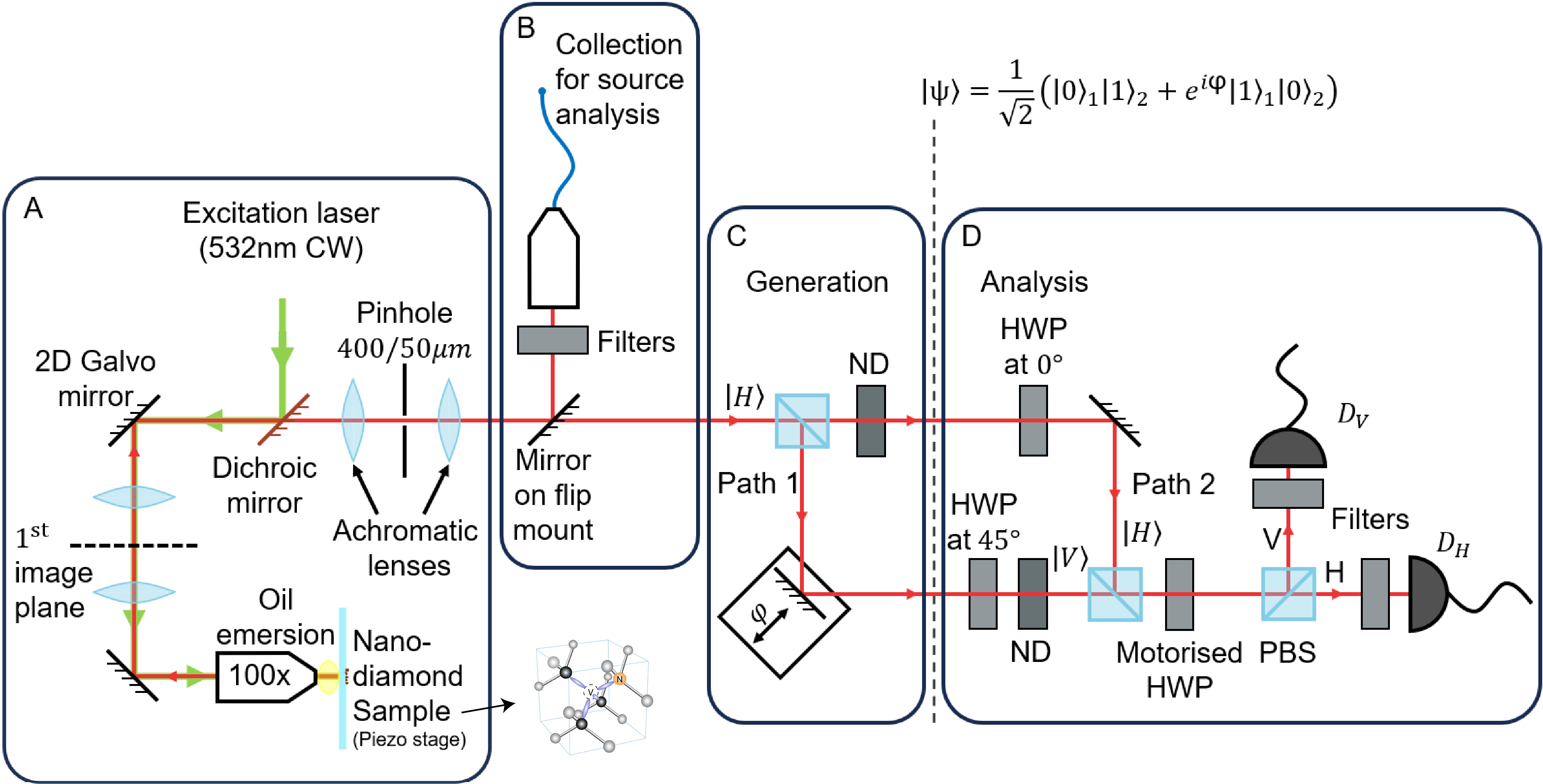}
    \caption{The full setup used for generating and verifying single-photon path entanglement with an NV center. Section A is a laser scanning confocal microscope system used for locating single NV centers on a coverslip substrate and collecting the single photons emitted. A region of a diamond lattice containing an NV center is shown next to section A. Here, the black and gray spheres are carbon atoms with N labeling nitrogen and V for vacancy. Section B is used for characterization of the single photons. Section C is used for entanglement generation and section D is used for analysis. See main text for details of each section. ND - neutral density filter, HWP - half-wave plate, PBS - polarizing beamsplitter, D$_i$ - detector for mode $i$.}
    \label{fig:full}
\end{figure*}

In most NV experiments, a pulsed laser is used to excite a single NV center so that the spontaneous emission represents a single-photon source~\cite{Kur00,Ber15, Aha11}. When using pulsed laser excitation, the pulses act as a trigger and photons detected within a coincident window of a pulse are considered produced by the NV center. A similar pulsed excitation method was used previously in Ref.~\cite{Pap09} to generate path-entangled single photons from an ensemble of caesium atoms. In our work, we use an NV center and propose a variation of their method using CW laser excitation to generate single-photon path entanglement. In our method, an artificial `trigger signal' is used to impose a state-generation window. The state-generation window mimics the effect of the coincidence window in the pulsed case, however, unlike the pulsed case, all photons detected can be considered. The benefit of this approach is that the duration of the state-generation window can be adapted to find the most appropriate time duration for generating a desired amount of entanglement, with higher generation rates possible at the expense of the amount of entanglement. Furthermore, from a fundamental perspective this opens up the possibility to observe a quantum-classical transition of entanglement in the system.

The paper is organized as follows. In Sec. II, we introduce the experimental setup, where we briefly characterize the NV center as a single-photon source, including a single-photon detected 2D fluorescence scan, second-order correlation measurement, as well as lifetime and spectrum analysis. In Sec. III, we then study the generation of path entanglement using CW laser excitation, verifying the presence of entanglement using the concurrence. We discuss the performance of our state generation window method and study the quantum-classical transition of entanglement in the system. We conclude the paper in Sec. IV by mentioning the key results of our work and future directions.


\section{Experimental setup}
Fig.~\ref{fig:full} shows a diagram of the full setup used in our work. Each section of the setup is aligned individually using multiple methods, each of which are mentioned in the relevant subsections that follow. Section A of the setup shows the stage that generates the single photons. Section B is used to locate and characterize the NV center used as our single-photon source. Section C generates the entangled state $\ket{\Psi}$ shown. Section D is used for analysis and to verify the presence of entanglement. In the next subsections we describe the different parts of the setup.


\subsection{Generation}
To generate the single photons needed for producing entanglement a laser-scanning confocal microscope is used -- a standard technique in single-particle fluorescence experiments~\cite{Jer18}. The microscope is shown in section A of Fig.~\ref{fig:full} and makes use of an infinity-corrected, high numerical aperture, oil-immersion microscope objective ($100\times,\ NA = 1.3$, Olympus UPlanFLN). The nanodiamond sample is prepared by diluting and sonicating a 1mg/mL suspension of nanodiamonds in DI water (Adamas Nanotech) and spin coating it onto a high precision $\#1.5$ coverslip made from Schott D 263 M Glass (Thorlabs, CG15XH1). The spatially distributed nanodiamonds are on average 40nm in diameter and contain 1-4 NV centers with a NV$^-$
/NV$^0$ ratio of $\sim 0.7$. The sample is excited using a continuous wave $532\ \mbox{nm}$ laser (Thorlabs, CPS532) with a pump power of $\sim 100\ \mu\mbox{W}$.  
\begin{figure}[t]
    \includegraphics[width=0.4\textwidth]{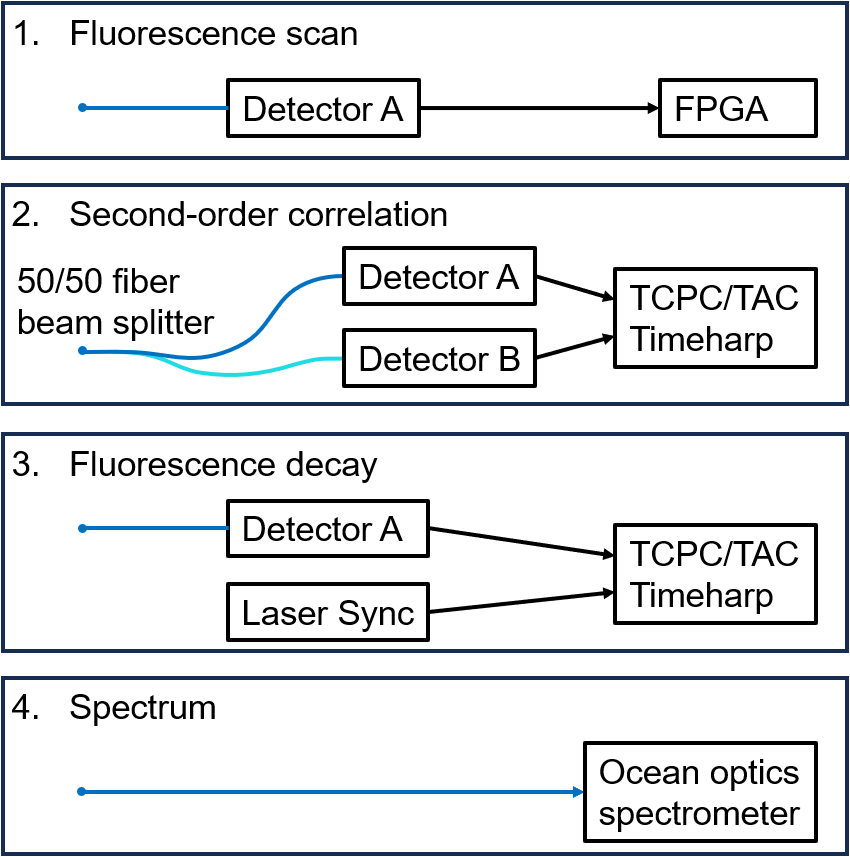}
    \caption{The different collection methods used for the different characterization steps of section B of the setup in Fig.~\ref{fig:full}. Box 1 is used for a fluorescence scan. Box 2 is used to do a $g^{(2)}$ measurement. Box 3 is used to measure the lifetime of the decay of the NV center. Box 4 is used to determine the spectrum of the photons emitted from the NV center. In all boxes, the signal from the detection equipment is sent to a PC for analysis.}
    \label{fig:analy}
\end{figure}

Light collected by the objective from different points on the sample surface exits the back aperture as collimated beams diverging from each other and represents an expanding image of the sample focused at infinity. The lens closest to the objective is placed one focal length from the back aperture and focuses each collimated beam individually while making the beams parallel to each other, creating an image plane ($1^{st}$ image plane in section A of Fig.~\ref{fig:full}, see Ref.~\cite{Fran21} for more details). 

A second lens is placed one focal length from the image plane and collimates each beam while making them converge to a point on a $2D$ galvonometric mirror (Thorlabs, GVS202). The lens directly after the $2D$ galvonometric mirror is placed one focal length from the point of convergence and focuses each beam while making them parallel to each other, creating a second image plane at the pinhole. The combination of the lenses closest to the galvonometric mirror turn the angular scanning of the mirror into a lateral translation across the pinhole. This results in the pinhole having conjugated points at both the first image plane as well as the sample. Therefore, only light from these points pass through the pinhole, filtering out neighboring points or points at different focal depths. The beam that passes through the pinhole is then collimated by the final lens which is placed one focal length from the second image plane. All lenses are achromatic doublets to reduce chromatic aberration and their placement at the focal lengths mentioned means that the magnification is not due to the lenses, but only due to the microscope objective. Further details of the setup can be found in Ref.~\cite{Fran21}

Using the generation section A of the setup an NV center photon source is located using a fluorescence scan and source characterization is performed by means of a second-order correlation measurement, as well as a lifetime and spectrum measurement using section B. We summarize briefly these different stages in the next section.


\subsection{Characterization}
Fig.~\ref{fig:analy} shows the four methods used for the analysis of the source. The measurements are a fluorescence scan, second-order correlation ($g^{(2)}$) measurement, fluorescence decay and a spectrum measurement. The measurements allow us to locate potential NV centers within nanodiamonds, determine if there is a single NV center present that acts as a single-photon emitter and identify what type of NV center it is.
\begin{figure}[t]
    \centering
    \includegraphics[width=0.35 \textwidth]{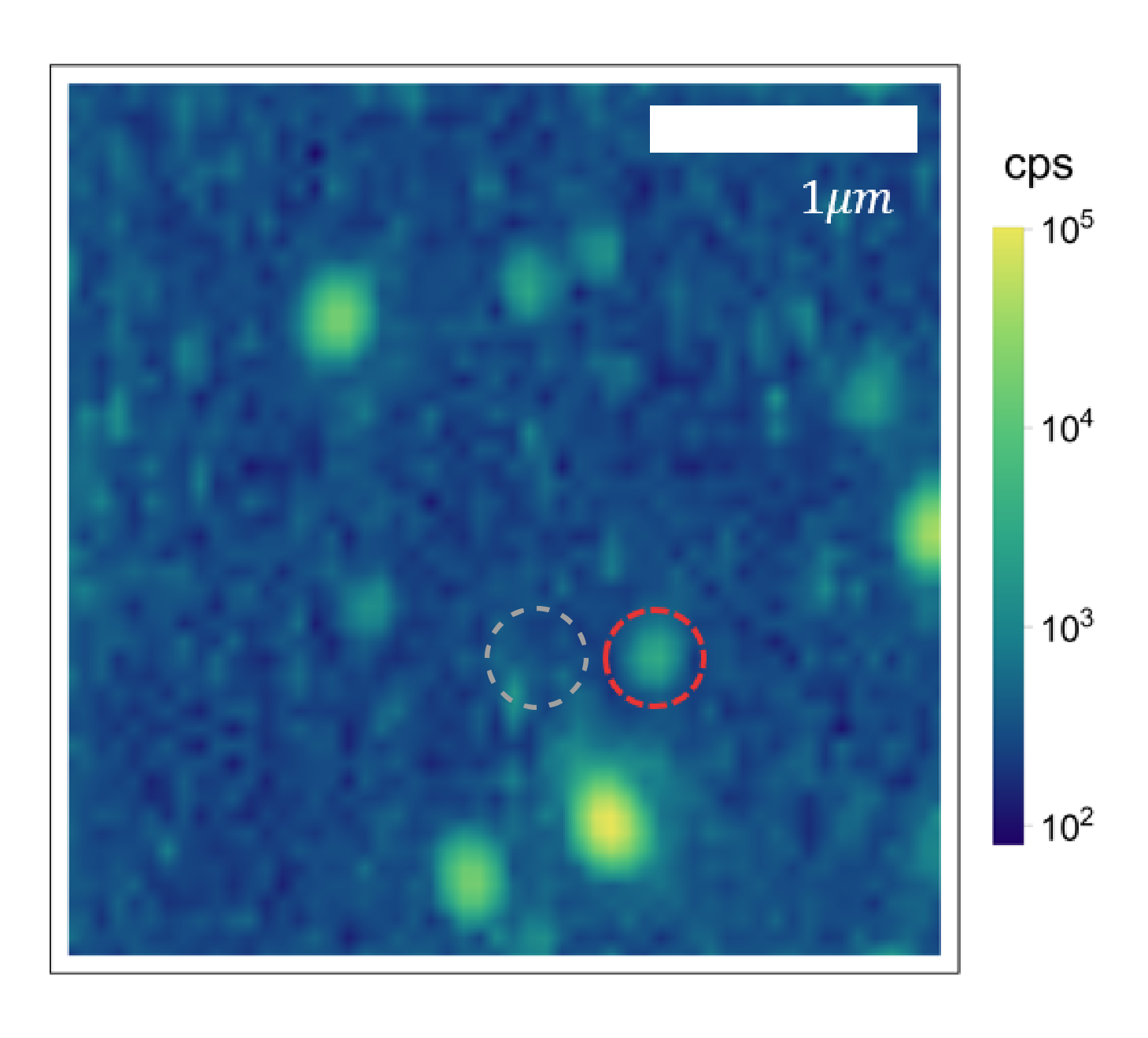}
    \caption{The intensity of the fluorescence over an area of the sample used to locate a single NV center. The red dashed circle indicates the area of the single NV center used for single-photon generation. The gray dashed circle shows the area that background photon measurements are taken (see later). Each pixel represents the photon counts over an area with integration time of $1 \mbox{ s}$ for each pixel.}
    \label{fig:scan}
\end{figure}


\subsubsection{Fluorescence scan}
A fluorescence scan is first performed to identify a nanodiamond with a single NV center that may be used as a single-photon source. The measurement is taken by using the $2D$ galvonometric mirror to move the excitation beam (and the collection spot) across an area of the sample while taking a measurement of the photon counts at each point on the sample using a single-photon avalanche diode (SPAD) detector (Excelitas, SPCM-AQRH-15-FC), labelled as detector A in box 1 of Fig.~\ref{fig:analy}. The collected light is first filtered using a 600\,nm high pass filter (Thorlabs, FELH600), a 800\,nm low pass filter (Thorlabs, FESH800) and a 10\,nm bandwidth notch filter at 533\,nm (Thorlabs, NF533-17). The 2D scan is represented as a density plot in Fig.~\ref{fig:scan}. Potential single-photon sources are identified by looking at the intensity of counts at different locations and identifying areas that have distinctly higher intensities than the background, but not high enough to indicate multiple NV centers being present. Once an area meeting these criteria is found and a photon source is identified the galvonometric mirror is shifted to align the point of excitation and fluorescence collection to where the source is present.

In order to ensure the photon source identified has only a single NV center emitter present a $g^{(2)}$ measurement is performed. In what follows we have used the bright spot in Fig.~\ref{fig:scan} with the red dashed ring. The detected collection area size $\delta r_d$ (lateral resolution in the $x$-$y$ plane) of our confocal microscope can be obtained from the formula $\delta r_d \approx \frac{d_d}{M}$, where $d_d$ is the pin hole diameter and $M$ is the magnification of the microscope objective~\cite{Bra90,Jer18}. In our case, we have used a $50$ $\mu\mbox{m}$ pinhole, giving $\delta r_d \approx 0.5\mbox{ }\mu \mbox{m}$ at the location of the dashed red ring in Fig.~\ref{fig:scan}.


\subsubsection{Second-order correlation}
\begin{figure}
    \includegraphics[width=0.45 \textwidth]{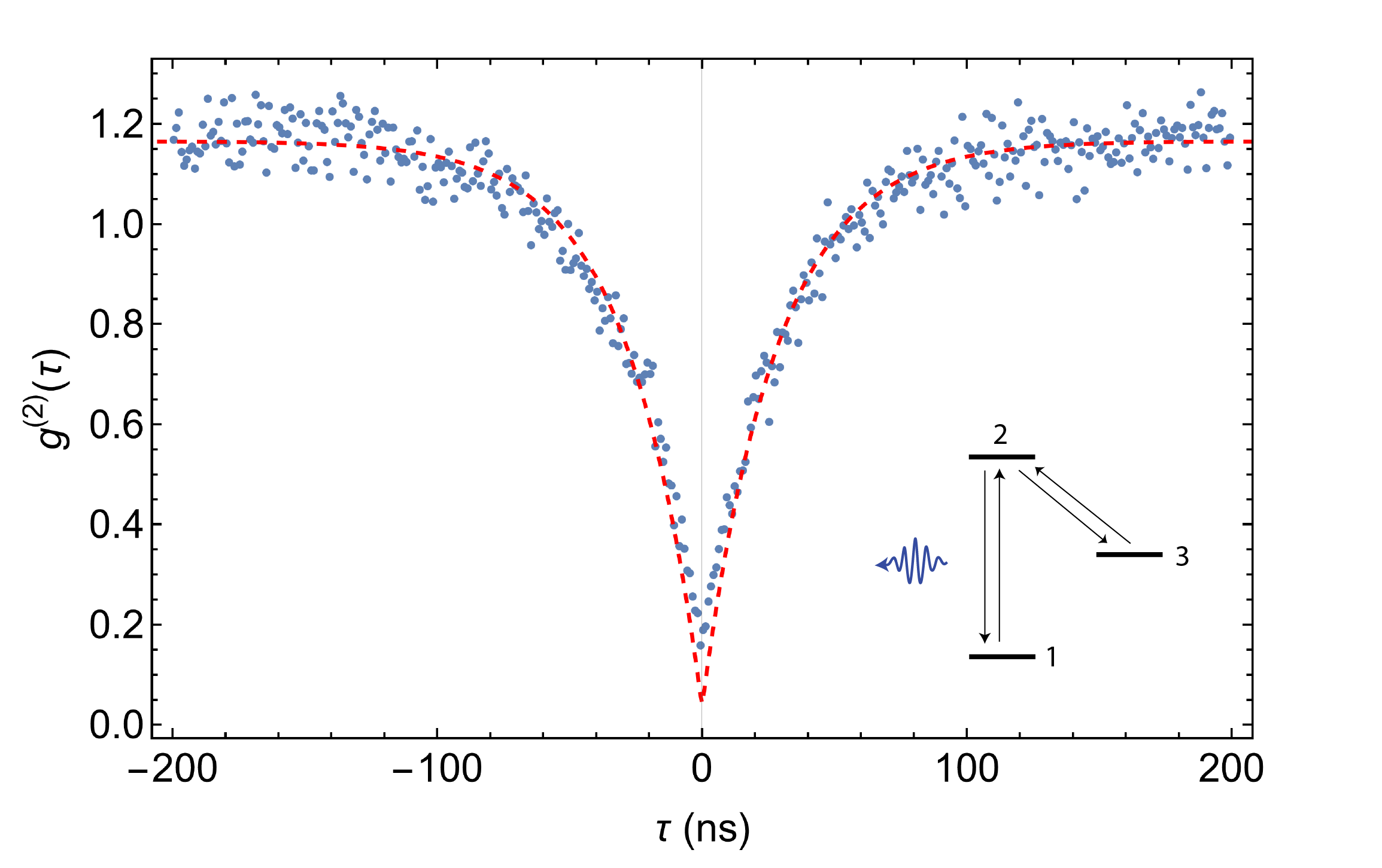}
    \caption{The second order-correlation function, $g^{(2)}(\tau)$. For ease of viewing the error bars on the data have been excluded. The red dashed line is a fit of the data to a theoretical model using a 3-level approximation for the NV center (inset shows the 3-level structure). The data is an average of five measurements taken over $1800\ \mbox{s}$ each with $\tau$ from $-200\ \mbox{ns}$ to $200\ \mbox{ns}$ in steps of $\Delta \tau = 1\ \mbox{ns}$.}
    \label{fig:g2}
\end{figure}

In Fig.~\ref{fig:g2} we provide a plot of the average of five measurements of the time-dependent second-order correlation, $g^{(2)}(\tau)$, taken using the photon source identified in the fluorescence scan. In this plot, the error bars are excluded for ease of viewing. Additionally, a theoretical model fit is shown as a red dashed line. This measurement allows us to determine whether the photon source identified using the fluorescence scan is a single-photon emitter, or if not then how many NV centers are present. 

To generate the plot shown in Fig.~\ref{fig:g2}, the $g^{(2)}$ measurement setup in box 2 of Fig.~\ref{fig:analy} is used, where both detectors are SPADs and a time-tagging unit (PicoQuant, TimeHarp 260 PICO) is used to measure photon arrival times at the two detectors A and B. The data is processed using the software QuCoa, which obtains the values using the equation
\begin{equation}
    g^{(2)}(\tau) =  c(\tau)/(N_1 N_2 w T).
    \label{eq:g2e}
\end{equation}
In Eq.~(\ref{eq:g2e}), $c(\tau)$ is the number of coincident counts within a time window $w$ with a time delay $\tau$ between detector A and B, $N_{1,2}$ is the number of photon counts in each detector, $T= 1800 \ \mbox{s}$ is the total integration time of the measurement, and $w = 1\ \mbox{ns}$ is the width of the coincidence time window. 

The $g^{(2)}$ measurement allows us to go beyond the basic photon counting from the fluorescence scan and verify temporal correlations in the statistics of the emitted photon state~\cite{Lou00}. A measurement of $g^{(2)}(\tau)$ is often interpreted as the probability of measuring a photon at time $t$ and another photon at time $t + \tau$~\cite{Ann15}. Here a multistart multistop time-to-digital conversion method is employed for the measurement of $g^{(2)}$ using the TimeHarp. Theoretically we expect $g^{(2)}(0) = 0$ for a single-photon source~\cite{Kim76}. For photon numbers equal to $n=1$, 2, 3, 4 and 5, we have that $g^{(2)}(0)=1-1/n$, which takes on the values $0,\ 0.5,\ 0.67,\ 0.75$ and $0.8$, respectively~\cite{Lou00}. From this we can conclude that for $g^{(2)}(0)<0.5$ the emitter is operating in the single-photon regime and by implication a single active NV center is present. As can be seen from the value of $g^{(2)}$ at zero-time delay in Fig.~\ref{fig:g2}, the NV center identified is clearly operating in the single-photon regime.
 
Theoretically, we expect $g^{(2)}(\tau)$ to have the following form~\cite{Ber15}
\begin{equation}
    g^{(2)}(\tau) = 1 - \beta e^{-\gamma_1 \abs{\tau}} + (\beta - 1) e^{- \gamma_2 \abs{\tau}},
    \label{eq:g2t}
\end{equation}
using a 3-level approximation of a single NV center (shown in the inset of Fig.~\ref{fig:g2}), with $\gamma_1,\ \gamma_2$ and $\beta$ related to rates within the NV center: $\gamma_1$ is the main decay rate while $\gamma_2$ is related to transfer into a metastable shelving state (shown as level 3 in the inset to Fig.~\ref{fig:g2} with $\gamma_2 \ll \gamma_1$) and leads to a small photon-bunching effect for NV$^0$ and larger for NV$^-$ (as the excitation power increases). This causes $g^{(2)}(\tau)$ to rise slightly above 1 on the wings of the curve, as seen in Fig.~\ref{fig:g2}. In addition, $\beta \simeq 1$ for NV$^0$, while it can grow larger for NV$^-$.

To take into account background scattering from the substrate we follow Ref.~\cite{Ber15} and include a factor $\rho$ to get $g^{(2)}_{fit}(\tau) = g^{(2)}(\tau) \rho^2 + 1 - \rho^2$ with $\rho= S/(S+B)$, where $S$ corresponds to the collected signal counts and $B$ the background counts for a fixed integration time. Using $g^{(2)}(\tau)$ from Eq.~\eqref{eq:g2t} we therefore have the following model to fit to our data
\begin{equation}
    g^{(2)}_{fit}(\tau) = 1 - \rho^2 \beta e^{- \gamma_1 \abs{\tau}} + \rho^2 (\beta - 1 ) e^{- \gamma_2 \abs{\tau}}.
    \label{eq:g2fit}
\end{equation}
We find $\beta = 1.18$, $\gamma_1 = 0.035$ ns$^{-1}$ and $\gamma_2 = 1.18 \times 10^{-4}$ ns$^{-1}$ using $\rho = 0.925$, where $S$ and $B$ are obtained by measuring counts at the red and gray dashed circles in Fig.~\ref{fig:scan}. The fit is shown as the red dashed line in Fig.~\ref{fig:g2}. Finally, from the data we find $g^{(2)}(0) = 0.173 \pm 0.039$. From this it can be concluded that the photon source identified using the fluorescence scan is a single-photon emitter and therefore a single NV center is present. 
\begin{figure}[t]
    \includegraphics[width=0.45 \textwidth]{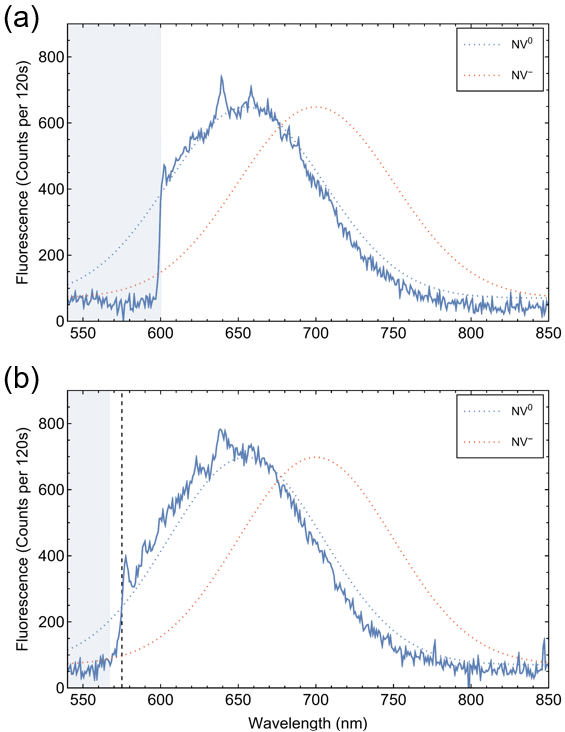}
    \caption{Average of 5 spectrum measurements of the NV center found with integration $120\ \mbox{s}$ for each measurement. (a) A $600\ \mbox{nm}$ high pass filter is used. (b) The $600\ \mbox{nm}$ high pass filter is removed so that the zero-phonon line is visible at $575\ \mbox{nm}$. In both cases the dichroic mirror used in the collection optics (see Fig.~\ref{fig:full}) excludes wavelengths below $567\ \mbox{nm}$.}
    \label{fig:spec}
\end{figure}

In Appendix~\ref{lifetime} we provide details of the measurement of the fluorescence decay lifetime of the NV center. We found that the decay constant is consistent with the value of $\gamma_1$ obtained from the second-order correlation model fit to our data. In the next section, we determine the type of NV center located, as the nanodiamond sample has a mix of NV$^-$ and NV$^0$ present.


\subsubsection{Spectrum}
In Fig.~\ref{fig:spec}~(a) we show an average of 5 measurements of the spectrum of the NV center taken over $120$ s using a spectrometer operating in the few-photon regime (Ocean Optics, QE Pro FL). The figure includes some visual guides to better identify the type of NV center present. It is expected that for an NV$^-$ (red dotted curve) the peak wavelength at room temperature should be $\sim 700\ \mbox{nm}$ with a ZPL at $637\ \mbox{nm}$. On the other hand, for an NV$^0$ (blue dotted curve) it is expected that the peak wavelength at room temperature is $\sim 655\ \mbox{nm}$ and the ZPL is at $575\ \mbox{nm}$~\cite{Aha11}. By inspecting the data shown in Fig.~\ref{fig:spec}~(a) it can be seen that the peak wavelength is $\sim 655\ \mbox{nm}$. From this it can be deduced that the NV center present is an NV$^0$. A measurement is also taken without the $600\ \mbox{nm}$ high pass filter to confirm the presence of the ZPL at $575\ \mbox{nm}$. In Fig.~\ref{fig:spec}~(b) the ZPL is visible at $575\ \mbox{nm}$ (black dashed line). From the data it is clear that the NV center found is a NV$^0$. We note that while removing the $600\ \mbox{nm}$ high pass filter increases the photon counts, the filter remains in place for the rest of our study to insure no back reflections from the pump beam interfere with the signal and filter out any potential Raman scattered light from the substrate~\cite{Ber15}. 


\section{Single-photon path entanglement}
Now that we have confirmed we have a single NV$^0$ acting as a single-photon source, we study the generation of entanglement using it. For entanglement generation, section C in Fig.~\ref{fig:full} is used, which consists of a non-polarizing beamsplitter and neutral density (ND) filter to balance the transmission and reflectance. Given a single photon input to section C, the state generated is ideally of the form 
\begin{equation}
\ket{\Psi} = \frac{1}{\sqrt{2}}\left( \ket{0}_1 \ket{1}_2 + e^{i \varphi} \ket{1}_1 \ket{0}_2 \right), 
\end{equation}
where the phase $\varphi$ is fixed via a mirror on a translation stage on path 1 and the polarization of the photon is set to horizontal ({\it i.e.} $\ket{1}_i = \ket{H}_i$) using a half-wave plate (HWP), quarter-wave plate (QWP) and polarizing beamsplitter (PBS) between sections B and C. All components used have broadband operation (400-800\ \mbox{nm}) so that we are able to send the photons equally into two output paths with a corresponding identical spectrum, which effectively decouples the spectral degree of freedom. The relative phase over the bandwidth of the NV emission is assumed to be approximately constant, an assumption that is validated by the visibility measurement described in section B. 


\subsection{Analysis and characterization}

For analysis and characterization of the state $\ket{\Psi}$, section D in Fig.~\ref{fig:full} is used. This section essentially completes a Mach-Zehnder (MZ) interferometer when combined with section C~\cite{Pap09}. When we perform the analysis in section D, the two paths from section C are made to be equal using a high-precision translation stage such that all wavelengths within the NV spectrum of each path can interfere and the broadband nature of the photons does not affect the degree of entanglement we measure. Several measurements are performed to verify that entanglement is generated in section C. The first of these measurements is a visibility measurement. The visibility $(V)$ of the analysis MZ interferometer quantifies the ability of a single photon, emitted from the NV center, to interfere with itself, thus giving a measure of the coherence of the terms in the state $\ket{\Psi}$. The visibility is calculated at an interference fringe using the equation
\begin{equation}
    V = \abs{P_{01}-P_{10}}.
    \label{eq:vis}
\end{equation}
In Eq.~(\ref{eq:vis}), $P_{01} = P_{H}$ and $P_{10} = P_{V}$ are normalized detection probabilities representing the probability of a single photon being detected in either $D_{H}$ or $D_{V}$. This path-to-polarization labeling comes from the photon in either path (1 or 2) being `tagged' using the orthogonal polarizations $H$ and $V$ in section~D: the polarization of the photon when in path 1 is rotated to vertical using a HWP, while when it is path 2 it remains horizontal. This effectively tags the path of the photon using the polarization. Thus, $P_V$ represents the probability the photon is in path 1 only ($P_{10}$) and $P_H$ the probability it is in path 2 only ($P_{01}$). The non-polarizing beamsplitter in section D simply combines the two paths (1 and 2) into the same output path, with an ND filter placed before it to balance the reflection and transmission. As the tagging is done with orthogonal polarizations there is no interference of the photon at the beamsplitter -- it simply acts as a path combiner (with a $50\%$ loss to the other output path).

The tagging process allows us to measure the ability of the two paths to interfere, {\it i.e.} the visibility, via an indirect measurement using the polarization degree of freedom in a single output mode. As the motorized HWP after the non-polarizing beamsplitter is rotated, the two paths are made to interfere indirectly through the polarization with which they were tagged. This allows the visibility to be measured using polarization measurements (see Ref.~\cite{Pap09} for details). Due to the interference the values for the normalized detection probabilities change as the photon interferes with itself through its polarization. The normalized detection probabilities are calculated using $P_i = N_i/(N_H + N_V)$ where $i$ represents either $H$ or $V$, and $N_i$ are photon counts. 

The visibility can be used with a quantity called the degree of contamination to verify whether entanglement is present in section C of Fig \ref{fig:full}. The degree of contamination, $y_c$, is calculated from the photon probabilities and quantifies the effect of two-photon contamination on entanglement~\cite{Pap09}. It is calculated using the photon probabilities $p_0$, $p_1$ and $p_2$, which represent the probability of zero, one and two photons (one in each path) being present. The values for the photon probabilities are obtained by assessing the number of photons detected during a fixed time window, which we call the `state-generation' window. The degree of contamination $y_c$ is then calculated using the equation
\begin{equation}
    y_c = 2 \left( \frac{N}{N-1} \right) \frac{p_2 p_0}{p_1^2},
    \label{eq:yc}
\end{equation}
where $N=2$ represents the number of modes present in the system~\cite{Pap09}. 

Using the calculated values of $V$ and $y_c$ the concurrence can then be calculated, which is used to verify the presence of entanglement. The concurrence, $C$, is a measure of bipartite entanglement generated in a system and is usually obtained directly by carrying out quantum state tomography and analysis of the density matrix~\cite{Schilling10, Israel12}. However, it is also possible by inspecting the density matrix to relate the concurrence to the visibility and degree of contamination~\cite{Chou05}, which are experimentally more accessible values to measure~\cite{Pap09,Lou09}. Given a state $\rho=p_0 \rho_0+p_1\rho_1+p_2\rho_2$, with at most one photon in either mode with probability $p=p_0+p_1+p_2$, the concurrence is given by $C={\rm max}((V-\sqrt{y_c})p_1/p,0)$ and provides a lower bound on the total entanglement in the system, $C_T \geq pC$~\cite{Chou05,Ray11}. The concurrence can be normalized by the one-photon probability to give the degree to which uncertainty bounds for a biseparable system are violated and therefore verify the presence of entanglement. The normalized concurrence is given by the equation~\cite{Pap09}
\begin{equation}
    C_N = \frac{pC}{p_1}={\rm max}(V - \sqrt{y_c},0).
    \label{eq:CN}
\end{equation}
This quantity can be viewed as the entanglement in the single-photon subspace: any product state present in the output of section C ({\it e.g.} a state resulting from classical light as a weak coherent state input) will always give $C_N=0$, as $y_C \geq 1$~\cite{Pap09}. This bound on classical or any biseparable states allows $C_N$ to be used a signature of entanglement and for a state with a vacuum part and a single-photon part, where the photon is entangled maximally in its single-photon subspace, we have $C_N=1$, as $V=1$ and $y_c\simeq 0$ (if $p_2$ is negligible compared to $p_0$ and $p_1$). Thus, in an ideal system with $V = 1$ and $y_c = 0$ we have $C_N = 1$, which represents a maximal violation of the uncertainty bound. On the other hand, in a system with no entanglement present we have $C_N = 0$, and for entanglement to be present we require that $C_N > 0$. In Eq.~(\ref{eq:CN}), it can be seen that for $C_N > 0$ we must have $V >\sqrt{y_c}$. This gives a lower bound on $V$ (for a fixed $y_c$) or an upper bound on $y_c$ (for a fixed $V$). 
\begin{figure}[t]
    \centering
    \includegraphics[width=0.4 \textwidth]{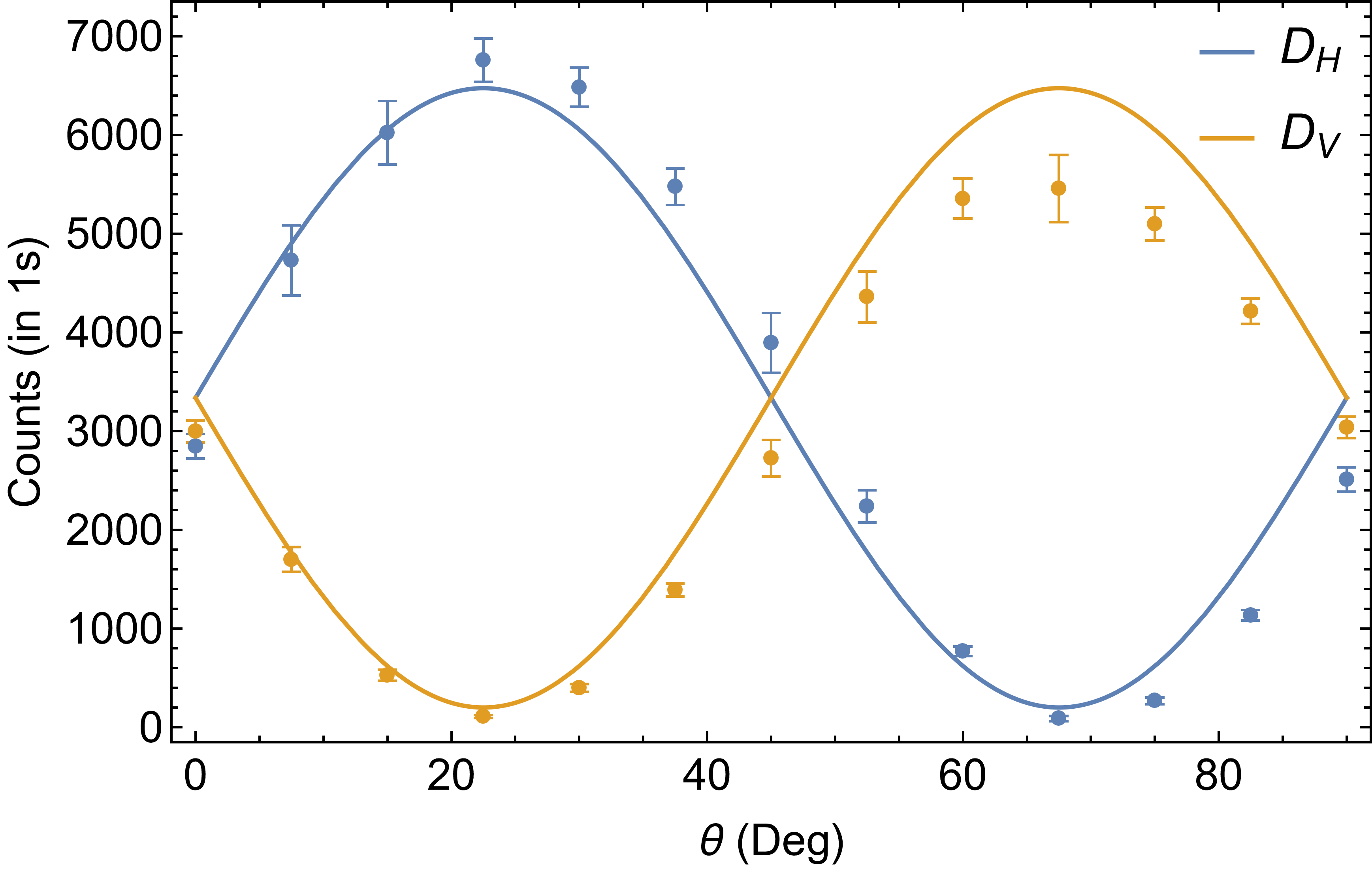}
    \caption{Visibility of the single-photon path entangled state. The plot shows the photon counts in $D_H$ and $D_V$ as a function of the angle of the motorized HWP $(\theta)$. $\theta$ is varied from $\theta = 0^\circ$ to $\theta = 90^\circ$ with $\Delta \theta = 2.5^\circ$, and an integration time of $20\ \mbox{s}$ for each point, giving a mean and standard deviation of 20 sets of $1 \mbox{ s}$ data at each point.}
    \label{fig:vis}
\end{figure}


\subsection{Visibility measurements}
We now discuss how the visibility is measured in our setup using photon counts measured in $D_H$ and $D_V$, which vary as the angle $(\theta)$ of the motorized HWP is changed. A motorized mount is used to reduce the amount of time taken to perform the measurement as the analysis interferometer in section D is unstable over long periods of time. When combining the two paths in the analysis section the state $\ket{\Psi}$ in the single-photon subspace is transformed into
\begin{equation}
    \ket{\Psi '} = \frac{1}{\sqrt{2}} \left( \ket{H} + e^{i \phi} \ket{V} \right),
    \label{eq:+}
\end{equation}
where $\phi$ is a sum of the phase gained inside the interferometer ($\varphi$) and the phase gained when combining the two paths into a single path using a non-polarizing beamsplitter, as the orthogonal polarizations pick up a relative phase. We set $\phi = 0$ by translating the mirror on path 1, which can be done as the components in the setup are roughly invariant over the spectral emission range of the NV center. When changing the angle $\theta$ of the fast axis of the motorized HWP relative to the vertical, the final state before the polarization measurement (using the polarizing beamsplitter and detectors $D_H$ and $D_V$) is given by 
\begin{equation}
    \ket{\Psi ''} = \frac{1}{\sqrt{2}}\Bigl(\bigl(\cos{2\theta} + \sin{2\theta}\bigr)\ket{H} + \bigl(\sin{2 \theta} - \cos{2 \theta}\bigr)\ket{V} \Bigr).
    \label{eq:visf}
\end{equation}
Thus, at $\theta = 0^{\circ}$ the counts in $D_H$ and $D_V$ should be equal as there is no interference of the paths. This can be seen clearly for our setup in Fig.~\ref{fig:vis}. At $\theta = 22.5^{\circ}$ we have that $D_H$ is maximized while $D_V$ should be minimized as the paths interfere maximally - Fig.~\ref{fig:vis} shows similar behaviour. At $\theta = 45^{\circ}$ we have that $D_H$ and $D_V$ are equal again as there is no interference. Fig.~\ref{fig:vis} shows this return to equal counts and the expected periodic nature of the interference as $\theta$ changes. A collective fit to the data is performed using the expected result from the first term of Eq.~(\ref{eq:visf}) and a variant for the second term (where the cosines and sines are exchanged), and is shown by the solid lines. 

The value for the visibility is calculated as the average of the visibilities at each interference fringe using the data from Fig.~\ref{fig:vis}. The mean visibility is found to be $V = 0.9329 \pm 0.0069$. This shows that the single photons from the NV center are able to interfere with themselves to a high degree, which is only possible if the path lengths are equal allowing the photons to interfere over the entire spectral range at the same time. Furthermore, such a high visibility suggests that the phase $\varphi$ of the path-entangled state over the bandwidth is approximately constant as differing phases would result in a large deviation of the visibility from unity.


\subsection{Population measurements}

Now that the visibility has been measured we determine $y_c$, and use both to obtain $C_N$ and verify the presence of entanglement. To determine $y_c$ a population measurement is performed. This is done by setting the HWP angle $\theta = 0$ in section D so that there is no interference and measuring photon events as they occur. The values of $p_0,\ p_1 \ \mbox{and }p_2$ are then obtained from the individual counts at the detectors and their coincidence counts via time-tagged measurements. As we are working with a maximum of $\sim$3,000 detected counts per second in any one detector (see Fig.~\ref{fig:vis}), the average time between counts is $\sim$330 $\mu$sec, which is much longer than the dead time of our detectors at 24~ns and so we do not expect the dead time to affect our measurements.
\begin{figure}[t]
    \centering
    \includegraphics[width=0.45 \textwidth]{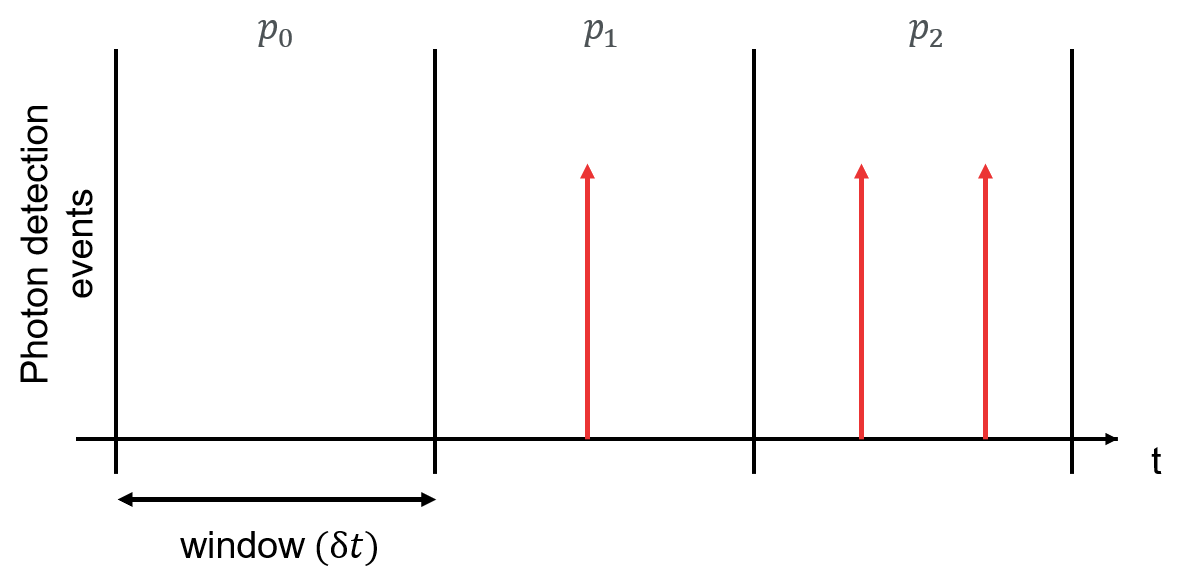}
    \caption{A visual depiction of how the state-generation window is used. Events where no photons are detected during a window of $\delta t$ count towards $p_0$, events where a photon is detected in either $D_H$ or $D_V$ during a window count towards $p_1$, and events where a photon is detected in both $D_H$ and $D_V$ within during a window count towards $p_2$.}
    \label{fig:stategen}
\end{figure}


\subsubsection{State-generation window}

In previous work~\cite{Pap09} the authors determined the populations of $p_0$, $p_1$ and $p_2$ when a pulsed laser was used as a sync signal. Photon events where the sync was not coincident with either $D_H$ or $D_V$, within a coincidence window, were then used to determine $p_0$. Additionally, photon events where the sync was coincident to either $D_H$ or $D_V$, within a coincidence window, were used to determine $p_1$, and photon events where the sync was coincident with both $D_H$ and $D_V$, within a coincidence window, were used to determine $p_2$. The coincident window therefore defined the period of time during which the state was considered, allowing a `generated state per pulse' to be considered. We have adapted the method of Ref.~\cite{Pap09} to function with a CW laser by imposing a time window, $\delta t$, that allows a `generated state per window' to be considered. This replaces the effect of the sync signal from the pulsed laser. In Fig.~\ref{fig:stategen} we show a diagram of the time windows in which we consider; no photons to be present, shown by $p_0$, one photon to be present, shown by $p_1$ and one photon in each path to be present, shown by $p_2$. Using this method the detected photon probabilities per state-generation window are determined. However, due to losses present within the system these detected photon probabilities are not the true photon probabilities of the generation stage of section C in Fig.~\ref{fig:full}. To determine the true photon probabilities the detection efficiency of section D is taken into account by including losses in the system.


\subsubsection{Losses}
To account for losses a measurement of the detection efficiency $(\eta_D)$ is taken and used to determine the true photon probabilities from the detected photon probabilities. This $\eta_D$ is not just the detection efficiency of the detector but rather a lumped detection efficiency which takes into account the attenuation of all components from the dashed line marked in Fig. \ref{fig:full} to the detectors. This is done using the set of equations~\cite{Pap09}
\begin{equation}
    \begin{split}
        p_2 = \frac{p^D_2}{\eta_D^2} ,\ \ \ 
        p_1 &= \frac{p^D_1 -2\eta_D\left( 1 - \eta_D \right) P^D_2}{\eta_D}\ \ \ \mbox{and}\\
        p_0 &= 1 - p_1 - p_2.
        \label{eq:loss}
    \end{split}
\end{equation}
Here, $p^D_i$ represents the detected photon probability and $p_i$ represents the true photon probability for photon number $i$.

To determine $\eta_D$, horizontally polarized laser light at 785~nm is sent into section D of the setup along path 2 and the intensity is measured in the path. The ND filter in path 1 before the beamsplitter is set to ensure that path 1 will have the same $\eta_D$ as path 2. The input intensity of the laser light in path 2 is $P = 80 \mu \mbox{W}$. A set of ND filters, with total transmission efficiency $\eta_{filters}= (4.59 \pm 0.79) \times 10^{-9}$, are then placed in front of $D_H$ to reduce the intensity of the light and avoid damage to the single-photon detector. Finally, a photon count measurement is taken and converted from $counts/s$ to $\mu$W such that the unit of measurement throughout is consistent, leading to a detected power $P_{D} = (1.477 \pm 0.0026) \times 10^{-8} \mu\mbox{W}$. These values are used to calculate $\eta_D$ using the equation
\begin{equation}
    \eta_D = \frac{P_{D}}{P \eta_{filters}},
    \label{eq:deteff}
\end{equation}
and yields a value of $\eta_D = 0.0402 \pm 0.0069$. The 785~nm laser light used to obtain this value is not the same as the peak emission of the NV center at 655~nm, however all components used are broadband (400-800~nm) and so the main difference is due to the detector efficiency (0.65 at 785~nm vs 0.68 at 655~nm)  according to the manufacturer's specifications which we account for in the detected power $P_D$. 
\begin{figure}[t]
    \raggedright
    \includegraphics[width=0.47 \textwidth]{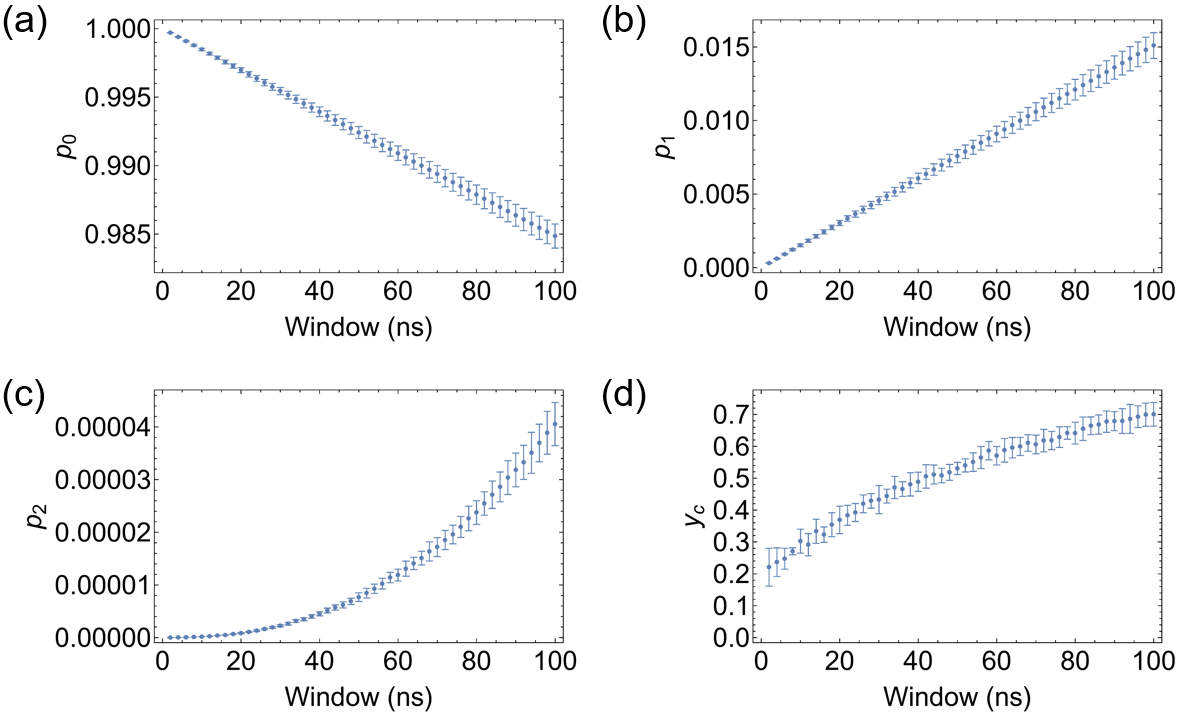}
    \caption{Populations and degree of contamination for the single-photon path-entangled state as a function of the size of the state-generation window. (a) Zero-photon probability, $p_0$. (b) One-photon probability, $p_1$. (c) Two-photon probability, $p_2$. (d) Degree of contamination, $y_c$. In each case, the state-generation window is varied between $2\ \mbox{ns}$ and $100\ \mbox{ns}$, with a step increase of $2\ \mbox{ns}$.}
    \label{fig:pop}
\end{figure}

Now that the detection efficiency is known the true photon populations can be determined and Eq.~(\ref{eq:loss}) used to calculate the degree of contamination.


\subsubsection{Photon probabilities and degree of contamination}
In Fig.~\ref{fig:pop} we show the photon probabilities $p_i$ and degree of contamination $y_c$ as the duration of the state-generation window is increased. For the photon probabilities it can be seen that as the size of the state-generation window increases $p_0$ decreases, while $p_1$ and $p_2$ increase. In particular, the increase in $p_2$ means that the larger the state-generation window the larger the two-photon contamination in the system. It can also be seen in Fig.~\ref{fig:pop} that as the size of the state-generation window increases the larger the errors present in the photon probabilities. This is due to the number of points sampled for each time window decreasing as the total data set size is fixed.

It can be seen in Fig.~\ref{fig:pop}~(d) that as the size of the state-generation window increases, $y_c$ increases. This is expected as the statistics of the light emitted from the NV center should tend to that of a Poisson distribution characteristic of classical light for a large coincident window, due to the behavior of $g^{(2)}(\tau)$ at large time delay~\cite{Lou00}. Thus, $y_c$ should tend to one and no entanglement should be present. On the other hand, a smaller size of the state-generation window reduces $y_c$ and ideally it should tend to zero, thereby maximizing the violation of the biseparability bound, with $C_N=1$. However, $y_c$ does not go to zero in our case as the NV center is not an ideal single-photon emitter due to its non-zero value of $g^{(2)}(\tau)$ for $\tau=0$, as seen in Fig.~\ref{fig:g2}. As a consistency check, in Appendix~\ref{g2pop} we calculate the populations $p_i$ and $y_c$ using $g^{(2)}(\tau)$ following the method outlined in Ref.~\cite{Stevens14} and find that the values match well. 

We note that the time tagging instrument we use, the TimeHarp 260 PICO, has a resolution of 25~ps, which means that in principle we can detect the arrival time of a photon to this resolution. However, the detectors have a jitter time of $\sim$350~ps, and so this is the lower limit of our resolution. The value is much smaller than the smallest time window we consider of 2~ns and so we do not expect timing synchronization to affect our measurements. 
\begin{figure}[t!]
    \centering
    \includegraphics[width=0.4 \textwidth]{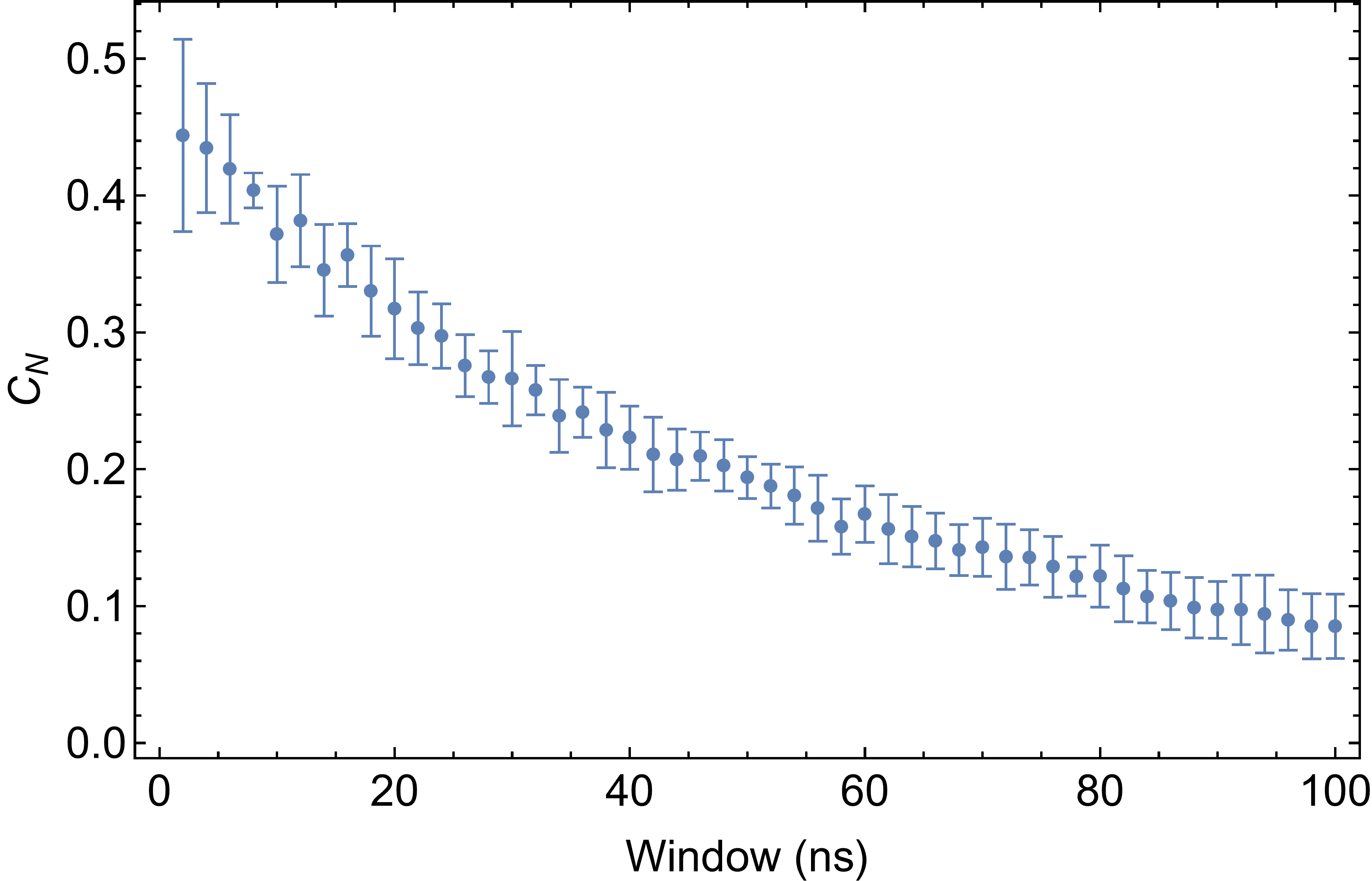}
    \caption{Concurrence, $C_N$, of the single-photon path-entangled state as a function of the size of the state-generation window. The window is varied between $2\ \mbox{ns}$ and $100\ \mbox{ns}$, with a step increase of $2\ \mbox{ns}$.}
    \label{fig:concur}
\end{figure}


\subsection{Entanglement}

The degree of contamination along with the visibility is now used to verify if entanglement is present in the generated state. In Fig.~\ref{fig:concur} we show the concurrence, $C_N$, as a function of the size of the state-generation window. It can be seen that as the size of the window decreases the concurrence $C_N$ increases. This is expected as ideally $C_N$ should tend to one as the size of the state-generation window tends to zero, as $y_c\rightarrow 0$. However, the trend toward one is not the case with our system as the concurrence is limited by the non-zero value of $y_c$, due to the non-zero value of $g^{(2)}(0)$. In addition, there is also a lower limit on the time window $\delta t = 2\mbox{ ns}$ considered -- at the count rates achieved we found 2~ns was the lowest window duration that gave enough coincidences to determine $p_2$ within a reasonable collection time, during which the analysis part of the setup was stable ($\sim 15$ min). Furthermore, the concurrence is also limited by the visibility achieved in our setup, although to a lesser extent due to the high visibility of $93$\%. The maximum value of the concurrence was found to be $C_N=0.44 \pm 0.07$ for a 2~ns window. 

As the size of the state-generation window increases, the concurrence $C_N$ can be seen to tend to zero in Fig.~\ref{fig:concur}. This should be expected because as the size of the window increases the statistics of the light tends to that of a Poisson distribution, behaving as classical light where entanglement cannot be present. However, $C_N > 0$ even for a large window of $100\ \mbox{ns}$. The long decay is likely due to the exclusion of additional two photon terms in the definition of the degree of contamination, where two photons are detected by the same detector (corresponding to 2 photons present in a given path), as well as the exclusion of three-photon and even four-photon terms, which are more likely at these large state-generation window sizes.

A very small state-generation window size ($<5$~ns) results in slightly larger error bars in $C_N$, which are likely due to the fact that for small window sizes the frequency of two-photon events is low and the relative error in $p_2$ increases. These larger error bars can be improved by increasing the sample size of the data while keeping the setup stable over the collection time required. Finally, we note that if all terms in the state $\rho$ emitted from the NV center are considered (not only the single-photon subspace), the total concurrence, $C$, is small due to the low value of $p_1$ compared to $p_0$ ($C=p_1 C_N \simeq 10^{-3}$ for a 20 ns window compared to $10^{-1}$ reported in Ref.~\cite{Pap09}). Improvements to the collection optics in our setup and the use of a solid immersion lens~\cite{Riedel14} or cavity-like structure~\cite{Huang19,Park23} that increases the capture efficiency of the emission would enable an increase in $p_1$, thus boosting the total concurrence. Despite this, the values obtained for $C_N$ are comparable to the values in Ref.~\cite{Pap09}. 


\section{Summary}

In this work we used a single-photon emitter and studied its ability to generate single-photon path entanglement. The emitter was characterized and identified to be an isolated nitrogen vacancy center, NV$^0$. Using the emitter with CW laser excitation we verified the presence of entanglement, giving a maximum value for the normalized concurrence of $C_N =0.44$, which is comparable to that obtained in Ref.~\cite{Pap09}. The benefits of our approach are that the emitter is a more accessible solid-state system operating at room temperature and that only CW excitation is needed. We also investigated the quantum-to-classical transition of the entanglement as the size of the state-generation window was increased.

Our setup could potentially be extended further to test fundamental features of quantum mechanics~\cite{Aspect89,Galvez05,Wang22,Aguilar24}, and verification and validation protocols for path-entangled states~\cite{Zhang21,Lachman19,Chunnilall14,Thomay17}. In the former case, it would be interesting to use our setup to verify the absence of entanglement at the output of a beamsplitter with a classical weak coherent state as input, providing further evidence that our analysis stage is working correctly. In the latter case of verification and validation, the measurement of the normalized concurrence requires a measurement of the visibility representing the wave nature of a photon and a measurement of the degree of contamination representing the particle nature. These measurements and their combination can be seen as a complementary approach to quantifying the quality of single photons from a quantum light source, {\it i.e.} quantifying the ability of a generated single photon to interfere with itself and become entangled.

Future work on improving our setup includes better filtering of the NV center emission to reduce the bandwidth of the photons and potentially increase the visibility~\cite{Galvez05}. While this would reduce the overall count rate, the collection efficiency could be improved to compensate, for instance, by using a solid immersion lens~\cite{Riedel14,Huang19,Park23}. This would also enable the state generation window to be reduced below 2 ns, reducing the degree of contamination and increasing the amount of entanglement. In terms of quantum applications, quantum communication protocols can be realized using single-photon path entangled states~\cite{Bjork12,Cas20} and the creation of a quantum communication network~\cite{Kur00}. Additionally, the number of paths may be increased, and the entanglement generated and utilized directly on-chip~\cite{Gao23}.


\section{Acknowledgements}
We thank Jason Francis and Alex Huck for their help developing the microscope stage. This research was supported by the Department of Science and Innovation (DSI) through the South African Quantum Technology Initiative (SA QuTI), Stellenbosch University (SU), the National Research Foundation (NRF), and the Council for Scientific and Industrial Research (CSIR).


\appendix
\renewcommand\thefigure{\thesection.\arabic{figure}}

\section{Fluorescence decay lifetime}\label{lifetime}
\setcounter{figure}{0}

To measure the fluorescence decay lifetime the NV center is excited using a $532\ \mbox{nm}$ pulsed laser with a pulse rate of $23.8 \mbox{MHz}$ (NKT Photonics, Fianium WhiteLase Micro), along with the fluorescence decay setup in box 3 of Fig.~\ref{fig:analy} used at the collection point in section B of Fig.~\ref{fig:full}. Here, the detector is a SPAD and a PicoQuant TimeHarp is used to measure the photon arrival times in relation to the laser pulse sync signal.
 \begin{figure}[t]
    \includegraphics[width=0.45 \textwidth]{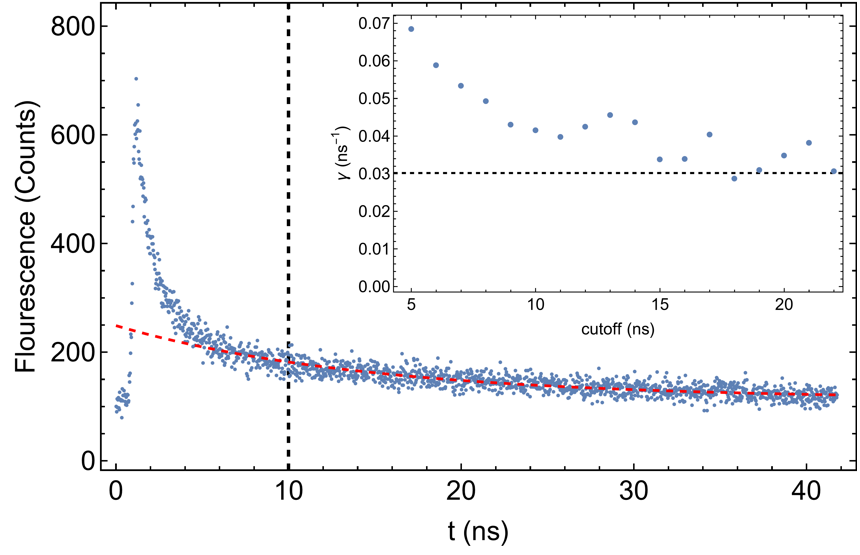}
    \caption{The decay of fluorescence from the NV center as a function of time, obtained using a pulse repetition rate of $23.8$ MHz and $\Delta t = 25$ ps, with an integration time of $20$ s. The black dashed line represents the cutoff time used to exclude the effect of the instrument response function. The red dashed line is a fit of the data to a theoretical model after the cutoff time. The inset shows the value of $\gamma$ obtained from the fit as a function of the cutoff time.}
    \label{fig:life}
\end{figure}

In Fig.~\ref{fig:life} we show the fluorescence signal of the NV center for increasing time from the laser pulse signal. This graph shows the decay of the fluorescence over time, as well as a theoretical fit shown in red and a cutoff point (dashed black vertical line) before which data is excluded from the fit to avoid the peak close to zero due to the instrument response time~\cite{Kur00}. The inset shows the effect of changing the cutoff point on the decay rate obtained from the fit. The value of the decay rate $\gamma_1$ from the $g^{(2)}$ fit is marked as a black dotted horizontal line. The theoretical fit in Fig.~\ref{fig:life} is obtained using the equation~\cite{Ber15}
\begin{equation}
    F(t) = \alpha e^{-\gamma t} + \beta,
    \label{eq:life}
\end{equation}
where $F(t)$ is the fluorescence intensity over time, $\alpha = 118.8 \pm 2.5$ is a normalization factor, $\beta = 97.8 \pm 5.3$ is a background correction factor and $\gamma = 0.0415\mbox{ ns}^{-1} \pm 0.0051\mbox{ ns}^{-1}$ is the decay constant.

When comparing the values of the decay constant shown in the inset of Fig.~\ref{fig:life} to that of $g^{(2)}$ it can be seen that as more of the initial peak is excluded the value of the decay constant approaches that obtained through the $g^{(2)}$ measurement. This is to be expected as the effect of the instrument response time is reduced when more data is excluded and shows consistency of our measurements.

\section{Second-order correlation and populations}\label{g2pop}
\setcounter{figure}{0}

Assuming $p_0 \gg p_1 \gg p_2$, the populations $p_0$, $p_1$ and $p_2$ can be obtained approximately from $g^{(2)}(0)$ for a fixed coincidence window $T$ and mean photon flux $f$. We calculate the populations this way as a consistency check. Within a coincidence window of length $T$ we have the mean photon number $\mu=fT$. From this we can write $p_2=g^{(2)}_D(0) \mu^2/2$, $p_1=\mu-2p_2$ and $p_0=1-p_1-p_2$, where $g^{(2)}_D(0)$ is the `detected' or measured value of $g^{(2)}(0)$ for a fixed $T$~\cite{Stevens14}. In the main text, we considered how the populations change over a range of $T$, requiring a knowledge of $g^{(2)}_D(0)$ for increasing $T$, which is not possible to obtain directly from the data.
\begin{figure}[b]
    \raggedright
    \includegraphics[width=0.52 \textwidth]{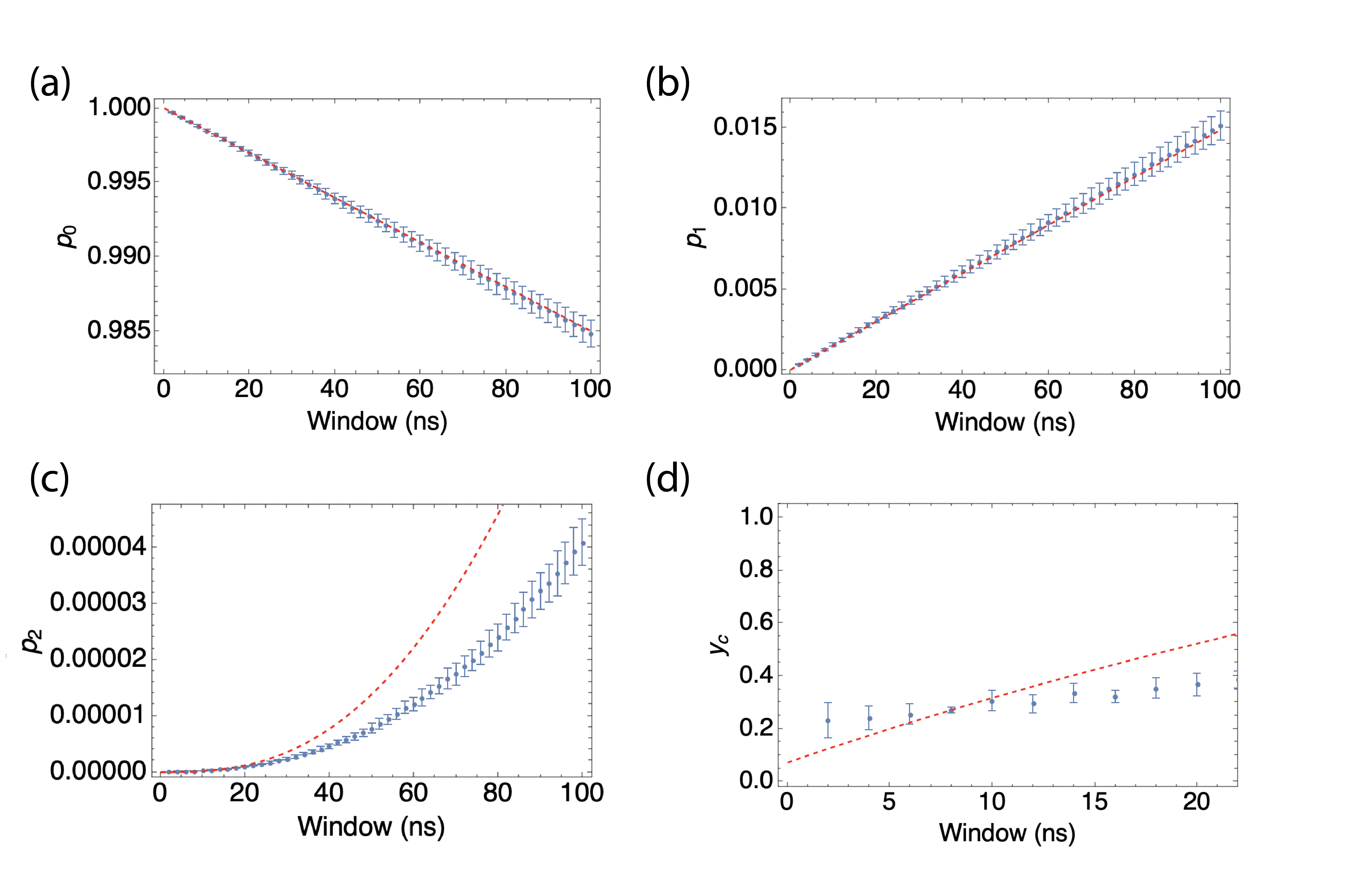}
    \caption{Populations and degree of contamination for the single-photon path-entangled state as a function of the size of the state-generation window, with dashed red lines corresponding to the values derived using the second-order correlation function. (a) Zero-photon probability, $p_0$. (b) One-photon probability, $p_1$. (c) Two-photon probability, $p_2$. (d) Degree of contamination, $y_c$ over the region where $p_2 \gg p_{n\ge 3}$.}
    \label{popfigapp}
\end{figure}
On the other hand, if the trend of $g^{(2)}(\tau)$ is known one can calculate $g^{(2)}_D(0)$ for an arbitrary coincidence window $T$ as follows. The detected value of the second-order correlation is defined as $g^{(2)}_D(0)=\langle m(m-1) \rangle_D/\langle m \rangle^2_D$, where for a stationary field~\cite{Lou00}
\be
\langle m(m-1) \rangle_D = \frac{2\mu^2}{T^2}\int_{t}^{t+T}dt' \int_{0}^{(t-t')+T}dt''' g^{(2)}(t'''), \label{g2det}
\ee
and $\langle m \rangle_D = \mu$. Thus with a knowledge of $g^{(2)}(\tau)$ one can calculate $g^{(2)}_D(0)$ for an arbitrary coincidence window $T$. For example, using the simple model $g^{(2)}(\tau)=1-e^{-\Gamma |\tau|}$ we find
\be
g^{(2)}_D(0) = \frac{1-e^{-\Gamma T}+\Gamma^2 T^2/2 -\Gamma T}{\Gamma^2T^2/2},
\ee
which is the well-known result given in Ref.~\cite{Lou00}. On the other hand, in our experiment we found the model fit~\cite{Ber15}
\begin{equation}
     g^{(2)}(\tau) = 1 - \beta e^{-\gamma_1 \abs{\tau}} + (\beta - 1) e^{- \gamma_2 \abs{\tau}},
    \label{eq:g2app}
\end{equation}
with $\beta = 1.18$, $\gamma_1 = 0.035$ ns$^{-1}$, $\gamma_2 = 1.18 \times 10^{-4}$ ns$^{-1}$ and $\rho = 0.925$. Using this model in Eq.~\eqref{g2det} we find
\bqa
g^{(2)}_D(0)&=&\frac{e^{-(\gamma_1+\gamma_2)T}}{T^2\gamma_1^2\gamma_2^2}\bigg(2\rho^2(e^{\gamma_1 T}(\beta-1)\gamma_1^2-e^{\gamma_2T}\beta \gamma_2^2) \nonumber \\
&&+e^{(\gamma_1+\gamma_2)T}(T^2\gamma_1^2\gamma_2^2+2(\beta \gamma_2^2-T \beta \gamma_1\gamma_2^2 \nonumber \\
&&+(\beta-1)\gamma_1^2(T\gamma_2-1))\rho^2)\bigg).
\eqa
The above formula is substituted into the relations for the populations, with $f=1.507 \times 10^{5}$ as the measured flux of photons in section C of the setup. The plots of the populations are shown in Fig.~\ref{popfigapp}. One can clearly see that the measured populations match very well those extracted using $g^{(2)}(\tau)$ for $p_0$ and $p_1$ (red dotted lines). On the other hand, in the case of $p_2$, there is a noticeable deviation of the measured populations compared to those obtained using $g^{(2)}(\tau)$. This is due to the presence of higher-order terms ($p_{n \ge 3}$) in the case where $g^{(2)}(\tau)$ is used, whereas the measured populations of $p_2$ in our experiment do not include these cases. Therefore, for large coincidence windows the presence of higher-order excitations should be included for a more precise value of the concurrence, although the values obtained are a lower bound as higher-order terms can in principle be filtered using local operations~\cite{Chou05}.


\end{document}